\documentclass[11pt]{article}
\usepackage{fullpage,amsthm}
\usepackage{amsmath,amssymb,amsfonts}
\usepackage{algorithmic}
\usepackage{graphicx}
\usepackage{textcomp}
\usepackage{xcolor}
\usepackage{lineno}
\usepackage{longtable}
\usepackage{indentfirst}
\usepackage{subcaption}
\usepackage{hyperref}

\begin{document}
\baselineskip 12pt

\begin{center}
\textbf{\Large Carruthers Data Processing Pipeline: \\ Instrument Effect Correction} \\

\vspace{1.5cc}
{ \sc Alex Zhang$^{*1}$, Heather Filippini$^{2}$, Lara Waldrop$^{1}$, Jason McPhate$^{3}$}\\

\vspace{0.3 cm}

{\small $^{1}$Department of Electrical and Computer Engineering, University of Illinois at Urbana-Champaign \\ 
$^{2}$Illinois Coordinated Sciences Laboratory, University of Illinois at Urbana-Champaign \\
$^{3}$Space Sciences Lab, University of California, Berkeley \\}
 \vspace{0.3 cm}
{\small $^{*}$Corresponding Author: alexmz2@illinois.edu}
 \end{center}

\vspace{1.5cc}

\begin{abstract}
  \noindent The Carruthers Geocorona Observatory—launched in September 2025—is NASA’s first mission devoted to investigating the fundamental nature of Earth’s exosphere from its distant vantage in halo orbit around the Earth-Sun L1 Lagrange point. Its primary payload, the GeoCoronal Imager, consists of two coaligned photometric imagers that measure ultraviolet Lyman-alpha emission radiance from exospheric hydrogen simultaneously at wide- and narrow- fields of view. The imagers use MicroChannel Plate intensified Complementary Metal Oxide Semiconductor detectors, which are known to add various artifacts to the final image telemetered from the spacecraft, hereby known as instrument effects. This paper details the algorithms used to retrieve and remove instrument effects from raw telemetry on-orbit, including detector voltage bias, thermal dark current, particle radiation, flat-field, and distortion. Finally, the science data processing pipeline from raw telemetry to instrument-effect corrected images is detailed. Algorithm performance is measured via a synthetic numerical image generator or validated on pre-launch experiments.

\vspace{0.95cc}
\parbox{24cc}{{\it Key words and Phrases}: Exosphere, Carruthers Geocorona Observatory, GCI instrument, Calibration
}
\end{abstract}

\section{Introduction}

The Carruthers Geocorona Observatory, launched in September 2025, is NASA's first mission dedicated to investigating the fundamental nature of Earth's exosphere. The mission aims to provide novel, high-cadence measurements from an unobstructed vantage point in a halo orbit around the Earth-Sun Lagrange-1 (L1) point. The primary target observable of the mission is the ultraviolet (UV) Lyman-alpha (Ly-$\alpha$) emission radiance at 1216{\AA}, which represents the brightest feature of Earth's geocorona \cite{meier_1991}.

To achieve the extreme sensitivity required to image the outermost extents of the exosphere, the mission's primary payload, the GeoCoronal Imager (GCI), employs MicroChannel Plate (MCP)-intensified Active Pixel Sensors (APS). While this electro-optical architecture enables the detection of faint, wide-field UV emissions, the physical detection and digitization processes inherently embed a variety of non-photon and hardware-induced artifacts into the raw telemetry, which are intrinsic to the instrument itself. Artifacts include electronic footprints like detector voltage bias and thermal dark current, as well as solar energetic particle radiation strikes on the detector, spatial distortion, and pixel-to-pixel gain variations across the MCP. We denote all of these artifacts as instrument effects. The precise retrieval and removal of these instrument effects is a strict prerequisite for the Carruthers mission's main science goal: performing high-fidelity exospheric science. This paper details the algorithmic framework designed to isolate and eliminate these hardware signatures from on-orbit data.

Table \ref{tab:data_product_def_table} defines the full hierarchy of Carruthers data products. This paper focuses strictly on the production of high-fidelity L1B images, which serve as the foundational input for subsequent science processing. The science data processing used to remove photon backgrounds was described in Zhang et al. (2026) \cite{Zhang26a}, while the absolute calibration algorithm was discussed in Zhang et al. (2026) \cite{Zhang26b}. The instrument effect retrieval and removal algorithms are validated either by pre-launch experiments \cite{Rider24cslgcisetup}, or by using a synthetic numerical image generator, which is described in Filippini et al. (2026) \cite{Filippini26}.

\begin{center}
\begin{longtable}{|c|c|c|} 
\caption{List of Data Product Definitions.} \label{tab:data_product_def_table}
 \\
\hline
Name & Description & Units \\
\hline
L0 & Raw images from spacecraft & [DN] \\
L1A & Dark rows and bias frame removed & [DN] \\
L1B & All instrument effects removed and dark corners zeroed out & [DN] \\
L1C & All photon backgrounds removed, cross-calibrated & $\left[\frac{\text{ photons}}{\text{cm}^2 \cdot \text{s}} \right]$ \\
L2 & 3D Hydrogen density & $\left[\frac{\text{atoms}}{\text{cm}^3} \right]$ \\
\hline
\end{longtable}
\end{center}

\subsection{Notation}

In this paper, upper-case letters (English or Greek) denote random variables, upper-case letters with an arrow, such as $\vec{X}$, denote random vectors, bold upper-case letters (English or Greek) denote matrices or sets, lower-case letters with an arrow, such as $\vec{x}$, denote vectors, and lower-case letters denote constants or indices. Variables $\epsilon$ that are a function of wavelength are denoted $\epsilon(\lambda)$, while variables $\epsilon$ that are a function of pixel position are denoted $\epsilon(i, j)$, where $i$ denotes the row index ($y$ direction) and $j$ denotes the column index ($x$ direction). Finally, all units are enclosed in brackets $[\cdot]$.

\section{Payload Description}

The GCI consists of two co-aligned imagers designed for the simultaneous sensing of exospheric Ly-$\alpha$. The Narrow-Field Imager (NFI) utilizes a 3.6$^{\circ}$ field of view (FOV) over $1024 \times 1024$ pixels to provide high spatial resolution near the Earth's limb, while the Wide-Field Imager (WFI) utilizes an 18$^{\circ}$ FOV over $512 \times 512$ pixels to capture the broader, dimmer extents of the exosphere. Despite their differing macroscopic views, both channels employ identical UV-intensified, APS-based cameras with heritage from the ICON mission's far-ultraviolet (FUV) imager \cite{mende2017iconfuvinstrument} and share a consistent, though physically independent, optical path as described below.

UV photons enter the instrument through an open entrance aperture and are reflected by Aluminum/MgF$_2$-coated curved mirrors that collimate the beam. The photons then pass through a 3mm-thick MgF$_2$ tube window featuring a 95\% transmissive Nickel (Ni) backside coating before striking a Potassium Bromide (KBr) photocathode. The detection of a photon by the cathode liberates an energetic photoelectron with a quantum efficiency of approximately 10\% \cite{carruthers_lab_cal_paper}. To support specific scientific objectives and wide-band spectral filtering, each channel further refines its baseline response using an independent 6-position filter wheel. One of the filters on this wheel is an aluminum blocking disk with zero transmissivity at all wavelengths. This filter is designed to block all incoming photons, which presents a convenient way to measure additive instrument effects.

The core of the instrument's detection capability, and the origin of the instrument effects addressed in this paper, lie in the signal amplification and digitization chain. Each primary photoelectron is accelerated across a $\sim$2100 V drop by a two-plate chevron MCP, where it triggers an electron avalanche \cite{darling2015performance}. This produces a localized charge cloud of roughly $10^5$ secondary electrons, which is then accelerated by a 3000 V potential onto a phosphor screen, converting the energy back into green visible-wavelength photons. These green photons are necked down through a fiberoptic taper onto the APS complementary metal oxide semiconductor (CMOS) detector.

Finally, these accumulated charges are read out by the camera's field programmable gate arrays (FPGAs) and digitized by an Analog-to-Digital Converter (ADC) to produce the final instrument units of Digital Numbers (DN). Every stage of this sequence introduces specific systemic artifacts: for example, the fused multifiber MCP bundles result in a spatially non-uniform gain, while the ADC introduces a distinct electronic bias in order to avoid reading negative electron counts. The algorithms presented in the following sections are specifically engineered to systematically decouple and remove these localized hardware effects from the resulting telemetry.

\section{Detector Voltage Bias}
\label{sec:bias_frame}

\begin{figure}
    \begin{subfigure}{0.48\textwidth}
      \includegraphics[width=\textwidth]{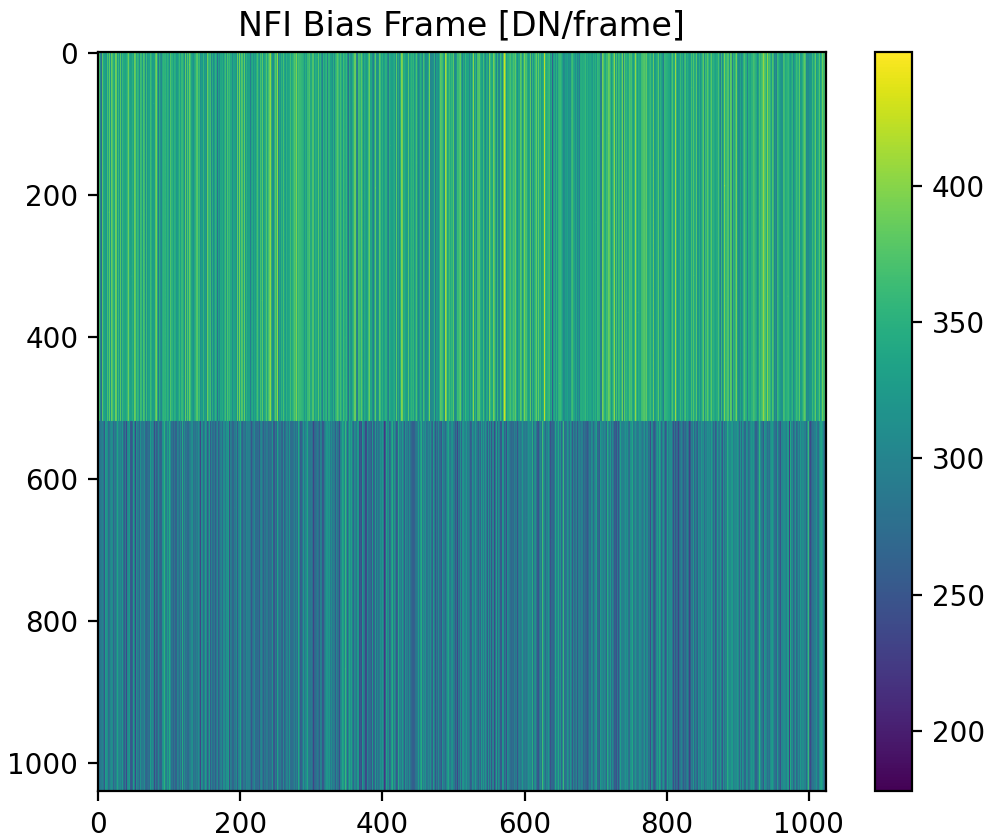}
      \caption{Example NFI Detector Voltage Bias}
      \label{fig:s3_lab_bias_frame_nfi}
    \end{subfigure}
    \hfill
    \begin{subfigure}{0.48\textwidth}
      \includegraphics[width=\textwidth]{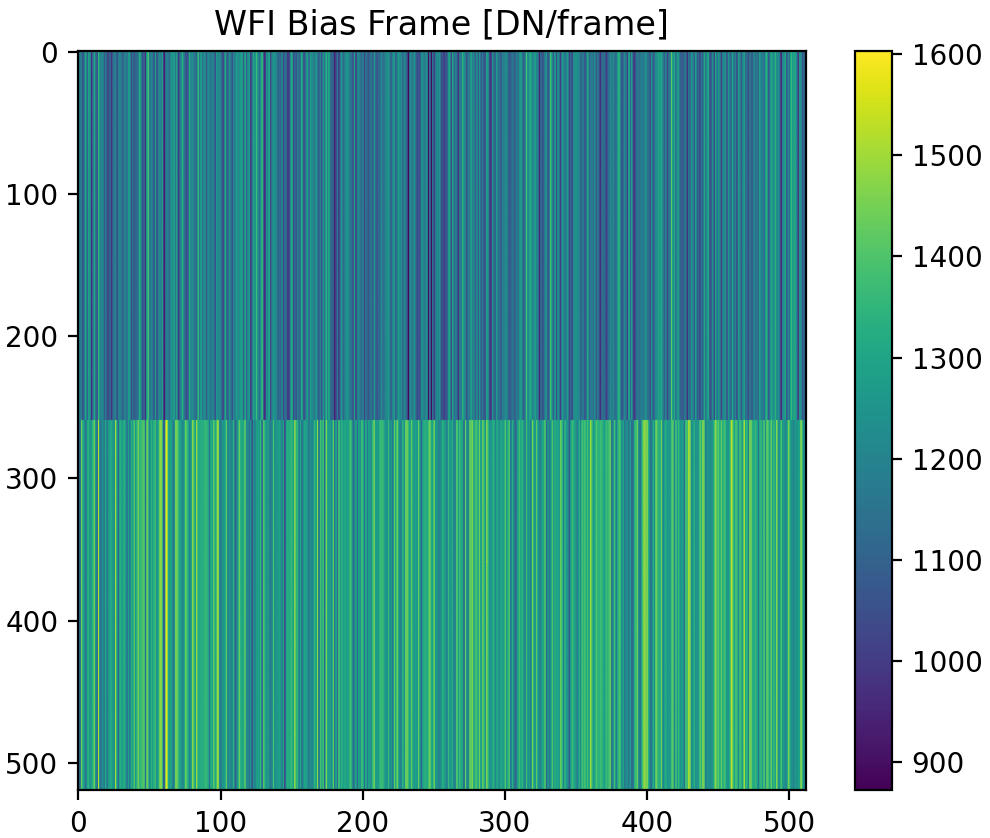}
      \caption{Example WFI Detector Voltage Bias}
      \label{fig:s3_lab_bias_frame_wfi}
    \end{subfigure}
    \caption{Representative detector voltage bias measurements derived from pre-launch laboratory tests. These frames were acquired using the blocked filter with the minimum integration time (65.6$\mu$s) at ambient temperature. The voltage bias applied to a binned WFI pixel is $\sim4$ times larger than that applied to a binned NFI pixel since WFI images are binned $4\times4$ before being telemetered, compared to the $2 \times 2$ binning for NFI. \label{fig:lab_bias_frame}}
\end{figure}

The electronic bias introduced by the ADC during readout constitutes a significant portion of the total signal and is the dominant feature in raw images. This voltage bias exhibits a distinct structure dictated by the detector electronics: since each APS is controlled by two separate FPGAs - one for the top half and one for the bottom - the voltage bias profile is split into two independent halves. Within each half, the voltage bias value is constant along vertical columns. As the two instrument channels utilize entirely separate FPGA banks, their detector voltage biases are also uncorrelated. Figure \ref{fig:lab_bias_frame} displays representative detector voltage bias frames acquired during pre-launch laboratory tests.

\subsection{Algorithm}
\label{sec:bias_frame_algorithm}

\begin{figure}[htbp]
    \begin{subfigure}[t]{0.48\textwidth}
      \includegraphics[width=\textwidth]{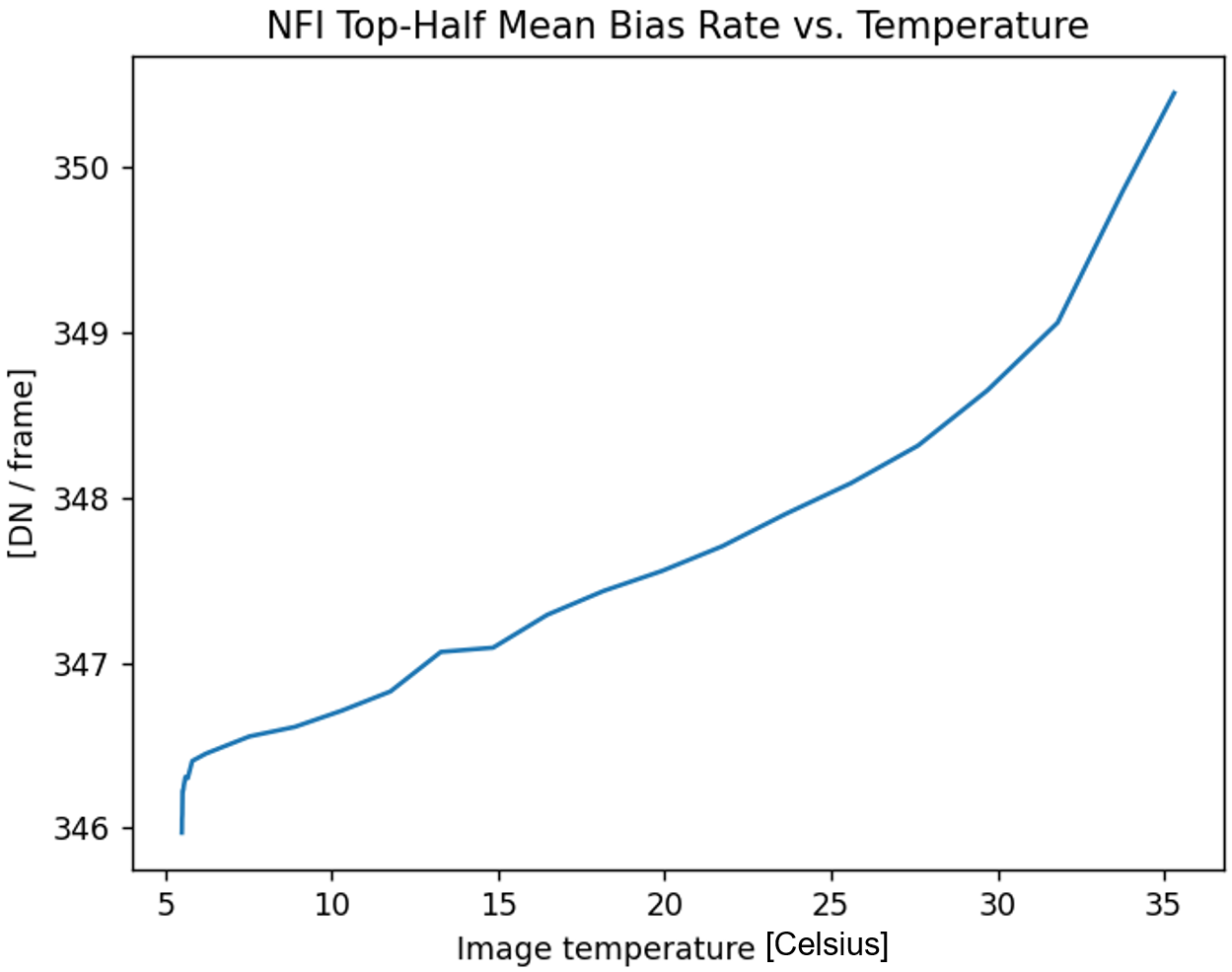}
      \caption{NFI top-half mean detector voltage bias [DN/frame] as a function of temperature.}
    \end{subfigure}
    \hfill
    \begin{subfigure}[t]{0.48\textwidth}
      \includegraphics[width=\textwidth]{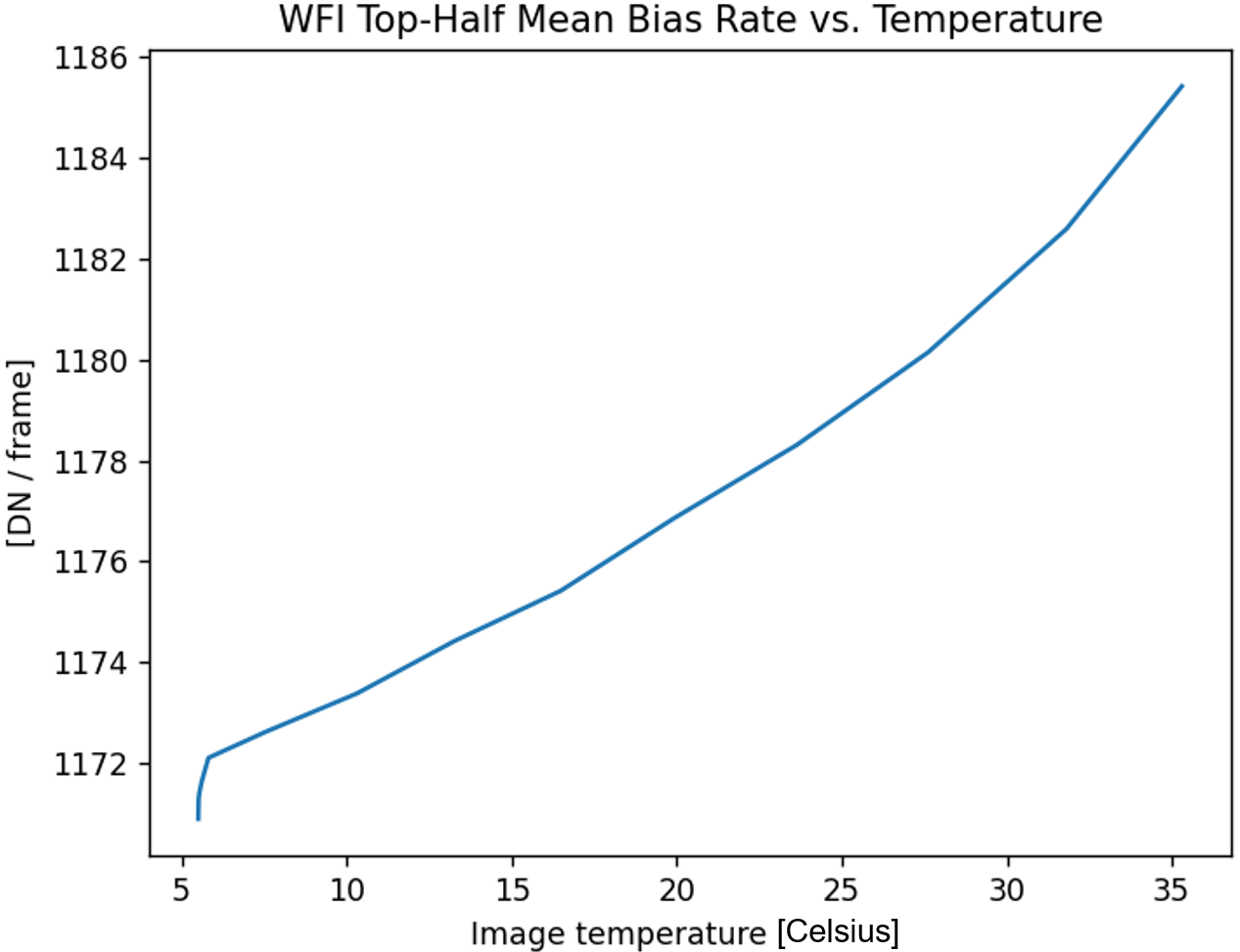}
      \caption{WFI top-half mean detector voltage bias [DN/frame] as a function of temperature.}
    \end{subfigure}
    \caption{Pre-launch laboratory characterization results. For acquisition specifications, refer to the caption in Figure \ref{fig:lab_bias_frame}. The bottom half of the detector voltage bias shows similar trends for both channels. \label{fig:bias_v_temp}}
\end{figure}

Although the detector bias can be measured directly on-orbit using full-frame, minimum-integration ($65.6\ \mu\text{s}$) images, this approach is ill-suited for the Carruthers mission. Because the bias voltage varies significantly with temperature (see Figure \ref{fig:bias_v_temp}), occasional calibration frames would fail to capture the instantaneous bias level contained in science images. Consequently, no dedicated bias frames are scheduled. Instead, the retrieval algorithm described below is employed.

Retrieval of the voltage bias independently for each image half is based on the signal measured in the electrically dark rows, which have their photodiodes disabled. These electrically dark rows are included with every frame read-out. After binning and frame stacking, each channel pixel in an electrically dark row $S_{\text{elec}}(p, q)$, in units of [DN], is modeled as:
\begin{equation*}
    S_{\text{elec}}(p, q) = \sum_{k = 1}^{n_{\text{frame}}} \sum_{i = pn_{\text{bin}}}^{(p+1)n_{\text{bin}} - 1} \sum_{j = qn_{\text{bin}}}^{(q+1)n_{\text{bin}} - 1} R(i,j,k) + b(i,j,k)
\end{equation*}
Here, $n_{\text{bin}}$ is the number of APS pixels binned into a single channel pixel in one dimension, with $n_{\text{bin}} = 2$ for NFI and $n_{\text{bin}} = 4$ for WFI. The number of image frames stacked to create the final raw image is denoted $n_{\text{frame}}$ (usually $480$ frames are stacked for each minute of integration). The detector voltage bias $b(i, j, k)$ is modeled as a constant with units of [DN], while read noise $R(i, j, k)$, which arises from the digitization process, is modeled as a zero-mean Gaussian random variable \cite{gow2007cmosreadout}.

Therefore, the voltage bias can be retrieved independently in every image by taking the mean of each half-image's electrically dark rows in the column direction (4 pixels for NFI, 2 pixels for WFI). The retrieved voltage bias is then normalized by $n_{\text{frame}}$ to obtain the bias signal per binned pixel, in units of [DN/frame], which is then archived by the Carruthers mission as the CAL\_BIAS calibration data product.

The bias removal algorithm is straightforward: scale the retrieved bias values in the CAL\_BIAS calibration data product by the number of frames in the relevant L0 image, then subtract the scaled value off of the L0 image.

Validation of the detector voltage bias retrieval algorithm on $100$ synthetic raw images for each channel with the shortest planned exposure (one-minute), generated using the numerical image simulator, demonstrated that the bias retrieval algorithm has a high-fidelity performance characterized by a negligible systematic error and a standard deviation of $<0.022\%$.

\section{Dark Current and Anomalous Pixels}
\label{sec:hot_pixels}

Thermal electron current due to the thermal motion of the electrons within each channel manifests as additional signal per second, or ``dark current''. Pre-launch laboratory measurements established an upper limit on the mean dark current of $1.25\ \text{[DN/s]}$ for NFI and $5.0\ \text{[DN/s]}$ for WFI. By comparison, the total incident UV photon signal (comprising both the exospheric target emission and the astrophysical photon background that the L1B data product is designed to isolate) ranges from approximately $100\ \text{[DN/s]}$ at the outermost edge of the exosphere up to $9000\ \text{[DN/s]}$ near the Earth's limb. Although dark current is modest relative to the brightest geocoronal emissions, it represents a non-negligible fraction of the faint outer-FOV photon signal and accumulates linearly with integration time. While the dark current rate depends on detector temperature, the sensor temperature on orbit is actively controlled and expected to vary slowly between 17$^{\circ}$C and 20$^{\circ}$C over the course of 5-6 hours. Across this narrow $3^{\circ}$C operational temperature range, pre-launch characterization indicated that variations in the mean dark current rate are negligible.

Anomalous pixels are defined as any pixel with abnormally bright or abnormally dim dark current. There are a host of reasons why these can appear, including fabrication variation in the CMOS device or aberrant leakage current on some component. Data collected during pre-launch laboratory tests indicate that most of these anomalous pixels are consistently bright, while a few anomalous pixels randomly and infrequently appear in one image and disappear in the next. The former can be identified and tracked in images taken with the blocked filter, which necessitates an anomalous pixel retrieval algorithm. Transient anomalous pixels cannot be corrected and will remain as artifacts in the final image; these artifacts are acceptable because they appear very infrequently.

Dark current and anomalous pixels are modeled as independent phenomena, but it is convenient to treat them together when removing both.

\subsection{Algorithm}

The anomalous pixel identification algorithm processes images taken with the blocked filter, after detector voltage bias has already been removed using the algorithm described in the previous section. Any pixel deviating from the global image mean by more than five standard deviations ($>5\sigma$) is flagged as anomalous. This detection threshold is a configurable parameter, allowing for on-orbit adjustments if necessary. The resulting coordinates are encoded in a binary mask and archived by the Carruthers mission as the CAL\_EPM (Extreme Pixel Map) calibration data product.

The retrieval algorithm was validated empirically during pre-launch laboratory calibrations. The number of anomalous pixels was measured to be around 2027 pixels for NFI (0.20\% of the total number of pixels) and around 618 pixels for WFI (0.23\% of the total number of pixels). The algorithm successfully identified all visually apparent anomalous pixels.

The dark current and anomalous pixel removal algorithm processes images by subtracting the nearest-in-time blocked (dark) exposure acquired under quiet radiation conditions. See Section \ref{sec:particle_radiation} for detailed radiation condition definitions. The quiet radiation condition is critical; subtracting a dark frame contaminated by high radiation would inadvertently remove signal required for subsequent processing steps. Operationally, a short blocked exposure is scheduled every three hours, ensuring that the reference dark frame is always temporally proximate. The algorithm is designed to prioritize the removal of anomalous pixels over precise dark current subtraction, accepting higher residual dark current error to ensure robust artifact removal.

\begin{figure}[htbp]
    \begin{subfigure}[t]{0.48\textwidth}
      \includegraphics[width=\textwidth]{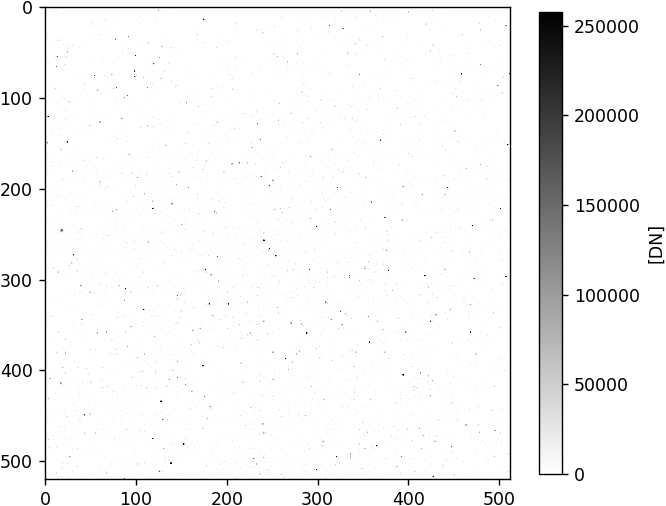}
      \caption{Before dark subtraction.}
      \label{fig:anom_pix_csl_val_pre}
    \end{subfigure}
    \hfill
    \begin{subfigure}[t]{0.48\textwidth}
      \includegraphics[width=\textwidth]{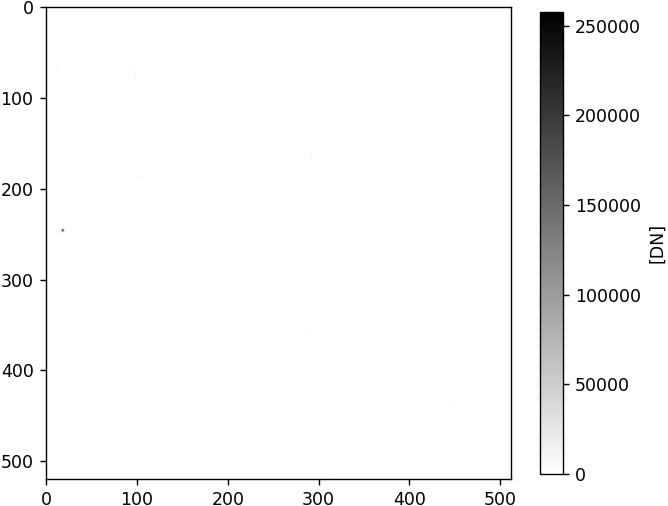}
      \caption{After dark subtraction.}
      \label{fig:anom_pix_csl_val_post}
    \end{subfigure}
    \caption{Comparison of a pre-launch laboratory L1A image (in digital numbers, DN) before (\subref{fig:anom_pix_csl_val_pre}) and after (\subref{fig:anom_pix_csl_val_post}) reference dark frame subtraction. Numerous anomalous pixels are distributed across the uncalibrated frame in (\subref{fig:anom_pix_csl_val_pre}). Dark subtraction effectively removes these artifacts in (\subref{fig:anom_pix_csl_val_post}), isolating the laboratory point source located near the left edge of the image.}
\end{figure}

Validation of the anomalous pixel removal algorithm relies on pre-launch laboratory data. Figure \ref{fig:anom_pix_csl_val_pre} displays an image (WFI channel, open filter) with the detector voltage bias already removed, which is dominated by localized point artifacts caused by anomalous pixels. Figure \ref{fig:anom_pix_csl_val_post} presents the same exposure following the subtraction of a blocked reference frame acquired 10 minutes prior. The subtraction successfully eliminates nearly all fixed-pattern point artifacts, leaving only the laboratory photon point source visible on the left. This result confirms that blocked-image subtraction is a highly effective strategy for mitigating anomalous pixels while preserving the true signal.

\section{Particle Radiation}
\label{sec:particle_radiation}

Any photoelectron or radiation ``event'' incident on the MCP, whether from photons or from energetic particles, undergoes gain amplification via an electron avalanche process. The number of electron counts per photoelectron event, denoted $G_{\text{mcp}}$ [counts/event], is controlled by the MCP voltage setting and does not follow any standard distribution. The amplification of each photoelectron event is assumed to be statistically independent. However, due to the MCP manufacturing process, the amplification efficiency is not spatially uniform across the detector face. Non-uniformities arise from stacking faults and channel distortions at the boundaries of the fused multifiber bundles, which locally alter the electron impact geometry and secondary emission efficiency. These pixel-to-pixel gain variations usually manifest as a fixed hexagonal pattern, which we will refer to as the MCP flat-field, $f_{\text{mcp}}(i, j)$. We define $f_{\text{mcp}}(i, j)$ such that it has a spatial mean of unity.

This section discusses detector backgrounds arising from energetic particle radiation at the Earth-Sun L1 orbit, rather than from UV photons transmitted through the optical system. At L1, the instrument is continuously exposed to Galactic Cosmic Rays (GCRs), which produce a persistent background of sparse, stochastic particle hits during nominal operations \cite{adams2013cremesim}. This background is periodically augmented by episodic Solar Energetic Particle (SEP) events, during which the energetic particle flux can increase substantially \cite{meier2014spaceweatherscale}. Although the ambient solar wind is also continuously present at L1, its predominantly low-energy particles are generally unable to penetrate typical spacecraft and instrument shielding and are therefore not considered a significant source of detector radiation events \cite{Kruk_WFIRST_radiation_background_on_Silicon}.

These backgrounds introduce two distinct detector components, each requiring a dedicated retrieval strategy. The first is MCP radiation, denoted $E_{\text{rad}}(k)$, caused by energetic charged particles depositing energy directly in the MCP. These events can originate from primary particles penetrating the instrument shielding or secondary charged particles generated through interactions with surrounding structures. Unlike photoelectrons generated by incident UV photons, particle-induced MCP events bypass the optical train and are therefore not subject to optical vignetting or wavelength-dependent optical throughput variations. However, they still undergo amplification by the MCP and are consequently modulated by the spatial MCP gain distribution. This provides a means of isolating and measuring the MCP flat-field term, $f_{\text{mcp}}(i, j)$. The second component is APS radiation, denoted $C_{\text{rad}}(k)$, which arises when energetic particles deposit energy in the silicon detector, generating spurious electron-hole pairs. In addition to direct particle interactions, secondary radiation generated in surrounding structures can contribute to this background. In particular, energetic electrons decelerating in surrounding materials can produce bremsstrahlung photons, which may subsequently reach and deposit energy in the silicon detector.

To quantify the expected radiation environment and establish model-based rates for our numerical image simulator, we utilized the CRÈME-MC engine \cite{adams2013cremesim} to simulate energetic particle interactions with the detector assembly behind 5mm of Aluminum (Al) shielding. For the APS, deposited energy in Silicon was converted to electron-hole pairs using an average energy of $3.6\ \text{eV/pair}$. Particle interactions with the instrument shielding and surrounding structures can also generate secondary particles and photons, including bremsstrahlung, which may contribute to the detector background. The radiation rates reported here are the model-derived rates used in the numerical image simulator; however, the present analysis does not provide a separate quantitative breakdown of direct primary-particle interactions and secondary radiation contributions. Accordingly, we do not assign a relative contribution to these mechanisms.

We evaluate the quiet-time radiation environment and three elevated SEP scenarios corresponding to NOAA's S2, S3, and S4 space weather levels, which are defined by energetic proton fluxes measured by the Geostationary Operational Environmental Satellites (GOES) \cite{meier2014spaceweatherscale}. The expected mean MCP radiation rates under quiet-time, S2, S3, and S4 conditions are $[0.075, 0.354, 3.53, 35.3]\ \text{[DN/s]}$ per binned pixel for NFI and $[0.299, 1.42, 14.1, 141]\ \text{[DN/s]}$ per binned pixel for WFI. Similarly, the expected APS radiation rates are $[0.006, 0.127, 1.27, 12.7]\ \text{[DN/s]}$ per binned pixel for NFI and $[0.025, 0.507, 5.07, 50.7]\ \text{[DN/s]}$ per binned pixel for WFI across the same conditions, respectively. Summing the MCP and APS contributions gives total expected radiation rates of approximately $[0.08, 0.48, 4.8, 48.0]\ \text{[DN/s]}$ per binned pixel for NFI and $[0.32, 1.93, 19.2, 191.7]\ \text{[DN/s]}$ per binned pixel for WFI.

For comparison, the UV photon-generated count rates that the L1B pipeline aims to isolate range from approximately $100\ \text{[DN/s]}$ at the outermost edge of the exosphere, where the interplanetary photon background dominates, to approximately $9000\ \text{[DN/s]}$ near the limb. Thus, under quiet-time conditions, the total radiation background is negligible, representing less than $0.4\%$ of even the lowest UV count rate for either instrument. Under S2 conditions, the contribution remains below approximately $2\%$ of the lowest UV count rate. However, during S3 SEP events, the radiation contribution increases to approximately $4.8\%$ of the lowest UV count rate for NFI and $19.2\%$ for WFI. During S4 events, these contributions increase to approximately $48\%$ and $192\%$, respectively. The radiation background is therefore generally negligible during nominal operations but can become a significant source of contamination during strong (S3) and severe (S4) SEP events, particularly for observations with relatively low UV count rates.

The temporal variability of these storm-time signals is well characterized by satellite monitoring. As demonstrated by observations of energetic proton fluxes across multiple solar radiation storms (e.g., Figure 1 of Mewaldt 2006 \cite{mewaldt2006sep_composition}), solar energetic particle intensities can evolve by several orders of magnitude over the course of a few hours. Given the Carruthers mission's 30-minute imaging cadence, consecutive exposures may therefore encounter substantially different radiation conditions. Consequently, the retrieval algorithms described below operate independently on individual images rather than assuming a temporally constant radiation background across multiple exposures.

\subsection{Algorithm: APS Radiation}

The APS radiation retrieval algorithm is applied to images following the removal of detector voltage bias, anomalous pixels, and dark current. In this corrected state, pixels located outside the FOV or in optically shielded rows are devoid of photon signal and MCP-amplified events; therefore, any residual signal in these regions is attributed exclusively to APS radiation. Because APS radiation consists of energetic particles penetrating the instrument housing to strike the silicon detector directly, the particle flux integrated over a multi-minute exposure is assumed to be spatially uniform across the entire focal plane. Furthermore, the CMOS detector responds identically to direct ionizing radiation in both the illuminated FOV and the unilluminated corners, as the FOV boundary is defined solely by the upstream MCP and optical geometry. The algorithm computes the mean intensity of these unilluminated reference pixels, normalized by the integration time, to yield the global APS radiation rate in units of [DN/s] to be subtracted across the full image. The resulting rate is archived by the Carruthers mission for each exposure as the CAL\_RAD calibration data product.

\subsection{Algorithm: MCP Radiation}

The MCP radiation retrieval algorithm is applied to images following the removal of detector voltage bias, anomalous pixels, dark current, and APS radiation. Unlike APS radiation, MCP radiation is spatially confined to the FOV, creating a direct degeneracy with the photon signal in non-blocked images. Consequently, the MCP radiation contribution can only be isolated in images acquired with the blocked filter, where there is no incident photon signal in the FOV. After removing detector voltage bias, anomalous pixels, dark current, and APS radiation, channel pixels in the FOV of these blocked exposures $S_{\text{blocked, fov}}(p, q)$, in [DN], can be modeled as
\begin{equation}
    \label{eq:rad_blocked_fov}
    S_{\text{blocked, fov}}(p, q) = \sum_{k = 1}^{n_{\text{frame}}} \sum_{i = pn_{\text{bin}}}^{(p+1)n_{\text{bin}} - 1} \sum_{j = qn_{\text{bin}}}^{(q+1)n_{\text{bin}} - 1} g_{\text{adc}} f_{\text{mcp}}(i, j) \sum_{n=1}^{E_{\text{rad}}(k)}G_{\text{mcp}}(n)
\end{equation}
Here, $n_{\text{bin}}$ is the number of APS pixels binned into a single channel pixel in one dimension, with $n_{\text{bin}} = 2$ for NFI and $n_{\text{bin}} = 4$ for WFI. The number of image frames stacked to create the final raw image is denoted $n_{\text{frame}}$. The constant $g_{\text{adc}} = 1/16$ represents the quantization step size of the digitization process in the ADC and is assumed to be known. Further details regarding this derivation are provided in Filippini et al. (2026) \cite{Filippini26}.

By definition, the spatial mean of the MCP flat-field $f_{\text{mcp}}(i, j)$ is unity across the FOV. Consequently, the MCP radiation rate is determined simply by computing the mean intensity of all FOV pixels in the blocked images and normalizing by the integration time.

However, a challenge arises because MCP radiation is present in all exposures, yet is directly measurable only in blocked images. To reconstruct the contribution from MCP radiation in non-blocked images, the algorithm utilizes a radiation scaling factor, $u_{\text{rad}}$, defined as the ratio between the APS and MCP radiation rates. The approach relies on the hypothesis that the two rates are strongly correlated, as both originate from the same flux of solar energetic particles. Since this assumption cannot be tested in a laboratory setting and lacks precedent in the literature, its validity must be assessed during on-orbit operations. Once established, $u_{\text{rad}}$ (derived from blocked images) allows the MCP contribution in non-blocked images to be estimated by scaling the concurrently measured APS signal.

For every blocked image, the measured MCP radiation rate and the calculated scaling factor $u_{\text{rad}}$ are archived by the Carruthers mission as part of the CAL\_RAD calibration data product.

\subsection{Algorithm: MCP Flat-Field}
\label{sec:mcp_flat_ret_algo}

The MCP flat-field retrieval algorithm is applied to images following the removal of detector voltage bias, anomalous pixels, dark current, and APS radiation. As noted in the previous section, MCP radiation is spatially confined to the FOV, creating a direct degeneracy with the photon signal. Consequently, the MCP flat-field can only be isolated in images acquired with the blocked filter, where the incident photon term $E_{\text{photon}}(i, j, k)$ is effectively zero. Equation \ref{eq:rad_blocked_fov} models the pixels inside the FOV. Given that the MCP flat-field $f_{\text{mcp}}(i, j)$ is defined to have a spatial mean of unity, the retrieval algorithm isolates this term by normalizing the image data by its global FOV mean. This operation effectively recovers the relative spatial gain distribution, $f_{\text{mcp}}(i, j)$. The resulting MCP flat-field is archived by the Carruthers mission as part of the CAL\_RAD calibration data product.

\subsection{Removal Algorithm: Radiation}

The radiation removal algorithm eliminates both APS and MCP signal contributions using parameters from the CAL\_RAD calibration data product. Applied to images with detector voltage bias, dark current, and anomalous pixels already removed, the algorithm first quantifies the ambient radiation environment by calculating the mean intensity of pixels outside the FOV. The algorithm then follows a conditional logic:
\begin{enumerate}
    \item Quiet-Time Regime: If the background mean is below a pre-specified threshold, the image is tagged as `quiet-time.' In this regime, contributions from any radiation source are assumed to be negligible, and no subtraction is performed.
    \item Active Regime: If the mean exceeds the threshold, then the algorithm retrieves and subtracts the uniform APS radiation component globally across the entire image. Subsequently, to remove MCP radiation (which cannot be directly measured in non-blocked images), the algorithm constructs an estimate using the current APS rate scaled by the radiation scaling factor ($u_{\text{rad}}$) and multiplied by the MCP flat-field ($f_{\text{mcp}}$) sourced from CAL\_RAD (measured from blocked exposures as described in Section \ref{sec:mcp_flat_ret_algo}, which directly captures the spatial gain modulation of particle hits on the MCP). This estimated MCP contribution is then subtracted from the FOV of the image.
\end{enumerate}

The algorithms in this section were validated by using the numerical image simulator to generate a quiet-time blocked image ($5$-minute integration time), a storm-time (S3) blocked image ($5$-minute integration time), a storm-time (S3) long blocked image ($30$-minute integration time), and a storm-time (S3) LyaN image on each channel. The LyaN filter was chosen because it is the current baseline science filter; the majority of the images taken on-orbit will use the LyaN filter. The NFI LyaN image had an integration time of $30$ minutes, while the WFI LyaN image had an integration time of $60$ minutes, which corresponds to the baseline image acquisition scheme for nadir-pointed science images on-orbit.

Simulated images were processed through the beginning of the calibration pipeline (voltage bias, dark current, and anomalous pixel removal). These images were then processed by the radiation retrieval algorithms described in this section. It was found that the retrieved signal from radiation exhibits a systematic underestimation of $\sim2.3\%$ across both channels, with standard deviations of $17\%$ (NFI) and $8.5\%$ (WFI). While these deviations appear significant, they must be contextualized by the signal magnitude: even in the S3 storm scenario, radiation constitutes only $\sim10\%$ of the total signal. Consequently, small absolute uncertainties translate into larger percentage errors for this component. Crucially, the percent error in the final science-filter images, after correcting for both radiation types, remains centered at $0\%$ with a standard deviation of just $1.3\%$ within the FOV.

\section{Flat-Field}

The GCI detectors exhibit inherent pixel-to-pixel sensitivity variations, collectively referred to as the flat-field, denoted $f(i, j)$. While standard calibration requires a uniformly illuminated source, no such target exists naturally in orbit. To generate an approximately uniform scene on-orbit, the spacecraft will slew off-nadir to target the InterPlanetary Hydrogen (IPH) emission, which is a diffuse, narrowband source chosen for its spatial linearity \cite{IPHModelPryor2013}. To eliminate residual spatial gradients in the IPH signal, the calibration sequence consists of four exposures acquired along a common boresight but separated by $90^{\circ}$ roll maneuvers. Due to the linear structure of the IPH background, stacking these rotated images effectively averages out the gradients, yielding an approximately uniform scene. This maneuver is scheduled every four weeks. All exposures utilize the open filter, based on the premise that the pixel-to-pixel non-uniformity is independent of filter.

The expected value of each pixel in the FOV for non-blocked images, denoted $\mathbb{E}\left[S_{\text{FOV}}(p, q)\right]$, was derived in Filippini et al. (2026) \cite{Filippini26}. Removing terms associated with voltage bias, APS radiation, dark current, anomalous pixels, and MCP radiation yields
\begin{equation*}
    \mathbb{E}\left[S_{\text{FOV, flat input}}(p, q)\right] = \sum_{k = 1}^{n_{\text{frame}}} \sum_{i=pn_{\text{bin}}}^ {(p+1)n_{\text{bin}} - 1} \sum_{j=qn_{\text{bin}}}^{(q+1)n_{\text{bin}} - 1} g_{\text{adc}}\mathbb{E}[G_{\text{mcp}}] f_{\text{mcp}}(i, j) f_{\text{opt}}(i, j) \iint \cdots dtd\lambda
\end{equation*}
Here, $n_{\text{bin}}$ is the number of APS pixels binned into a single channel pixel in one dimension, with $n_{\text{bin}} = 2$ for NFI and $n_{\text{bin}} = 4$ for WFI. The number of image frames stacked to create the final raw image is denoted $n_{\text{frame}}$. The constant $g_{\text{adc}} = 1/16$ represents the quantization step size of the digitization process in the ADC and is assumed to be known, while $\mathbb{E}[G_{\text{mcp}}]$ denotes the mean of the MCP gain, which is also assumed to be known. The argument in the integral is not relevant to this section and is thus omitted.

\begin{figure}[hbtp]
    \begin{subfigure}{0.48\textwidth}
      \includegraphics[width=\textwidth]{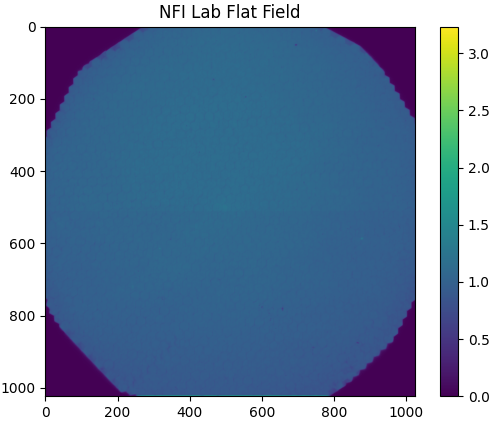}
      \caption{NFI}
      \label{fig:s7_lab_flat_field_nfi}
    \end{subfigure}
    \hfill
    \begin{subfigure}{0.48\textwidth}
      \includegraphics[width=\textwidth]{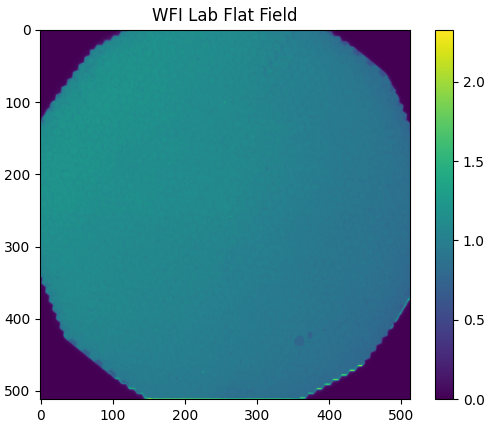}
      \caption{WFI}
      \label{fig:s7_lab_flat_field_wfi}
    \end{subfigure}
    \caption{Flat-field $f(i,j) = f_{\text{mcp}}(i,j) f_{\text{opt}}(i,j)$ for both channels as measured during pre-launch laboratory calibrations\label{fig:lab_flat_field}}
\end{figure}

The objective of the flat-field retrieval algorithm is to recover the composite flat-field, $f(i,j)$, which encapsulates both the MCP gain $f_{\text{mcp}}(i, j)$ and optical throughput $f_{\text{opt}}(i, j)$. The former was defined in Section \ref{sec:particle_radiation}, while the latter accounts for spatial variations in transmission efficiency such as vignetting that are not covered by the MCP gain flat-field. While Section \ref{sec:mcp_flat_ret_algo} presented a method for isolating the MCP flat-field, the MCP flat-field retrieval yielded insufficient precision. Consequently, the radiation-derived MCP flat-field is utilized exclusively for the removal of MCP radiation artifacts. For science data correction, a dedicated retrieval of the full flat-field $f(i,j)$ is performed here. The total flat-field is normalized such that its spatial mean across the FOV is exactly unity. See Figure \ref{fig:lab_flat_field} for pre-launch examples.

\subsection{Algorithm}

The flat-field retrieval algorithm processes images following the removal of detector voltage bias, anomalous pixels, dark current, and radiation artifacts. To isolate the detector response, bright celestial sources such as stars, the Moon, and the outer planets are identified using the algorithms described in Zhang et al. (2026) \cite{Zhang26a} and are masked in order to prevent high-intensity point sources from biasing the flat-field calculation. Crucially, the $90^{\circ}$ roll maneuver ensures that celestial sources shift position relative to the detector between exposures. Consequently, a pixel masked due to a bright object in one frame typically remains valid in the other three, ensuring continuous coverage of all pixels across the entire detector. To distinguish between true celestial sources and intrinsic detector features (which might mimic bright spots), the masking algorithm performs a persistence check: any pixel masked in all four exposures is re-classified as a static flat-field feature and subsequently unmasked.

To retrieve the flat-field, the four calibrated images are combined by computing the pixel-wise mean (excluding masked values). The resulting composite is normalized by the global FOV mean to enforce an average response of unity. The derived flat-field is archived by the Carruthers mission as the CAL\_FLAT calibration product. The flat-field is removed from science images by dividing pixels within the FOV by the nearest-in-time available flat-field.

Validation of the flat-field algorithms uses the numerical image simulator to generate a sequence of quiet-time flat-field images (30-minute image with the blocked filter followed by $4\times 90$-minute images with the open filter, each rotated $90^{\circ}$). Simulated L0 images were processed through the beginning of the calibration pipeline (voltage bias, dark current, anomalous pixel removal, radiation signals, celestial source mask). These images were then processed by the flat-field retrieval algorithm described in this section. For both channels, the percent error on the flat-field is found to be centered at $0\%$, with standard deviations of $3\%$ (NFI) and $1.43\%$ (WFI). Both the mean bias and the standard deviation fall well within $5\%$, which indicates that the retrieval algorithm meets the $5\%$ accuracy requirement mandated by NASA. While a negligible fraction of pixels ($\sim0.15\%$) exhibit outliers exceeding $10\%$ error, these are statistically insignificant and excluded from the primary analysis.

This algorithm has been further validated analytically using the instrument model developed in Filippini et al. (2026) \cite{Filippini26}. It can be shown that the mean-square error of the recovered flat-field in each pixel is no more than 5\%  of the ground-truth for both channels in all except the worst radiation storm conditions (S4 and above).

\section{Distortion}

Optical distortion, where the solid angle of sky subtended by a pixel varies across the focal plane, is present in both channels. The optical distortion is particularly noticeable in the WFI due to its wide $18^{\circ}$ FOV. Optical distortion maps were generated during pre-launch laboratory tests by translating a UV point source across a uniform grid in the laboratory frame. The deviation of the recorded spot positions from a uniform grid on the image plane quantified the optical distortion. On-orbit, images of dense star fields will be taken on both channels to measure the deviations between expected stellar positions and measured stellar positions. The algorithm to retrieve distortion coefficients will be the same as the one applied to the lab data, described by Sirk et al. (2026) \cite{carruthers_lab_cal_paper}.

\section{Registration}

\begin{figure}
    \begin{subfigure}{0.3\textwidth}
      \includegraphics[height=2.2in]{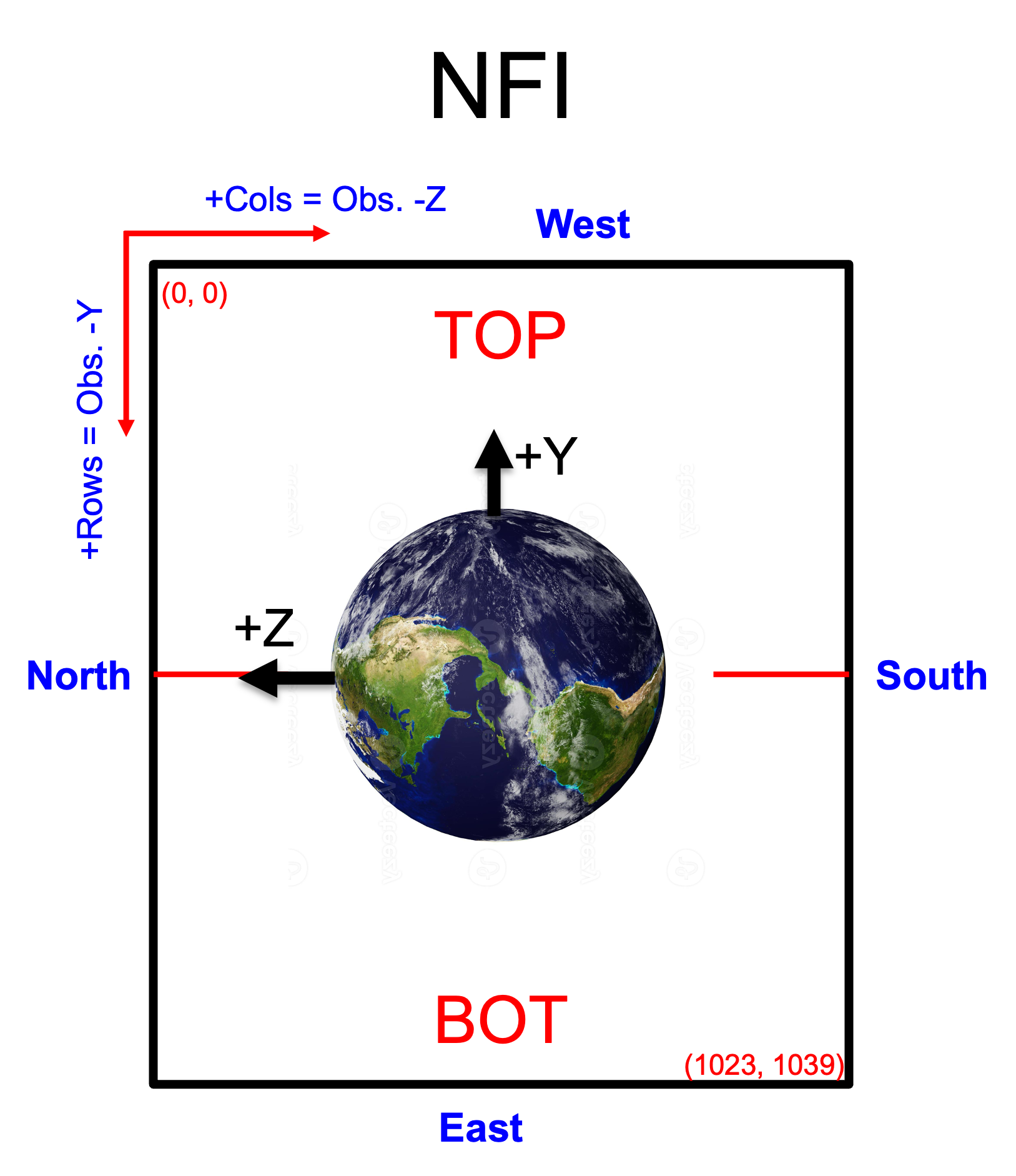}
      \caption{Projection of Earth and GSE y and z axes in NFI coordinates for the spacecraft positioned at L1.}
      \label{fig:NFI_L0_coordinates}
    \end{subfigure}
    \hfill
    \begin{subfigure}{0.3\textwidth}
      \includegraphics[height=2.2in]{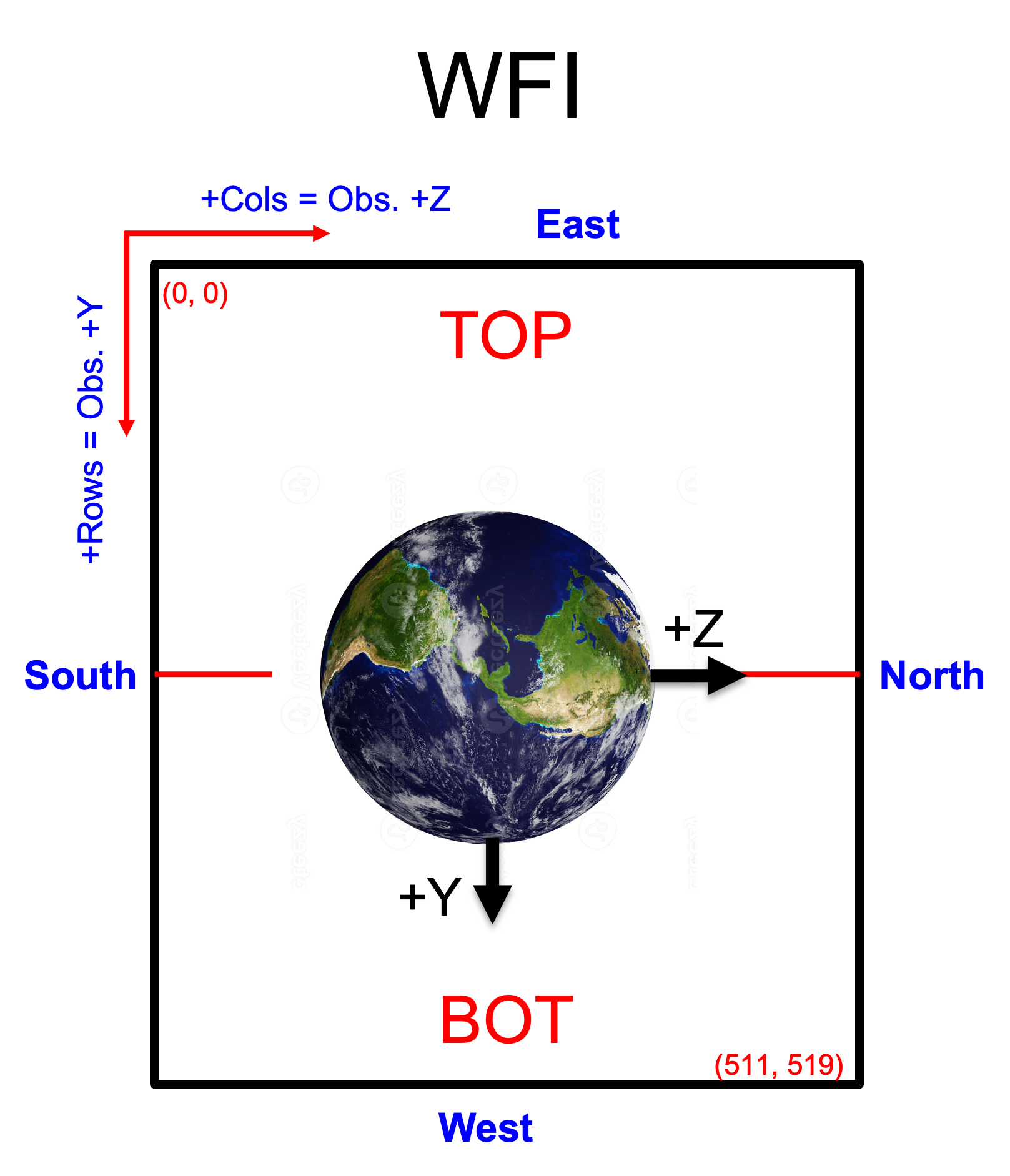}
      \caption{Projection of Earth and GSE y and z axes in WFI coordinates for the spacecraft positioned at L1.}
      \label{fig:WFI_L0_coordinates}
    \end{subfigure}
    \hfill
    \begin{subfigure}{0.3\textwidth}
      \includegraphics[height=2.2in]{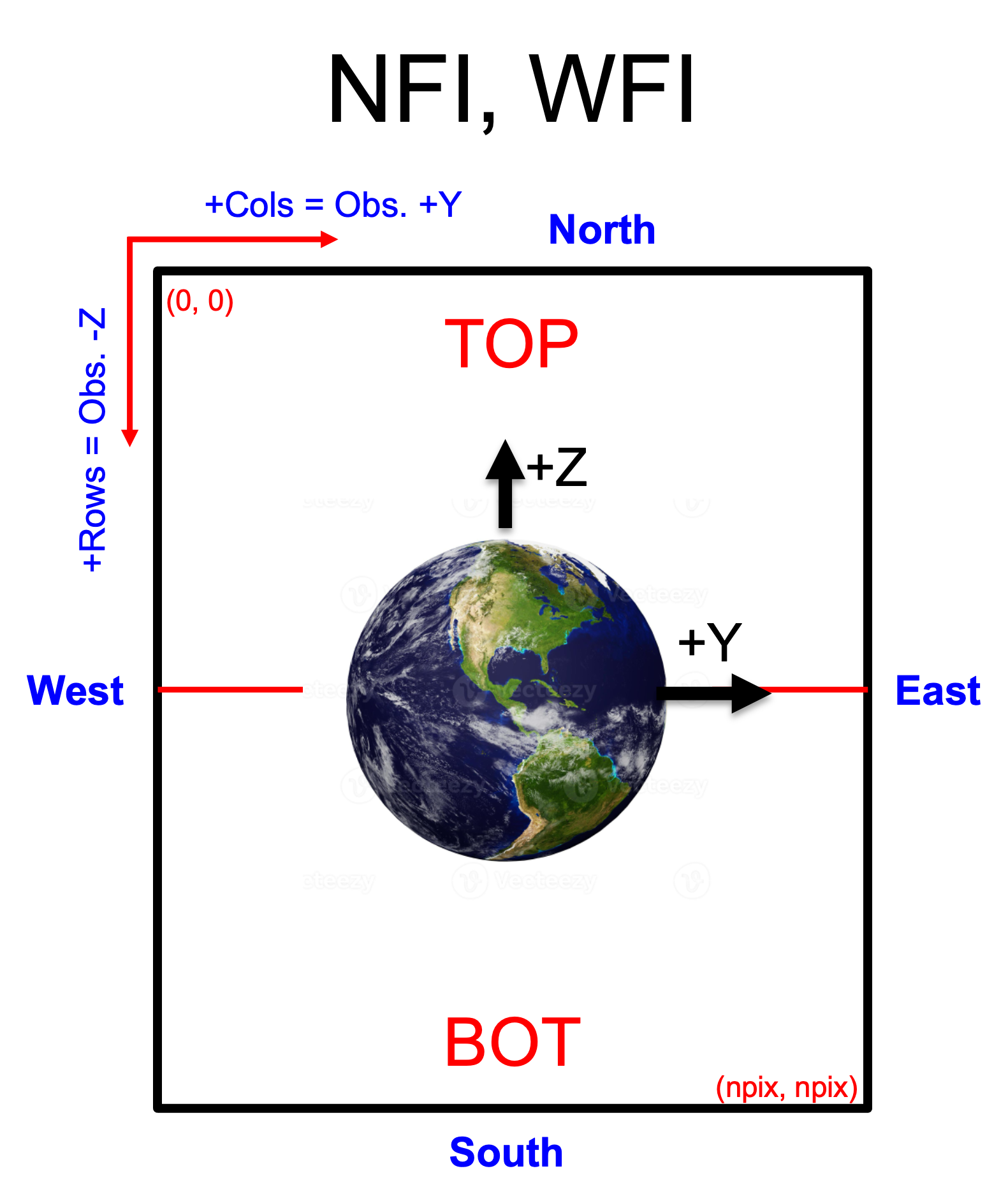}
      \caption{Projection of Earth and GSE y and z axes in registered coordinates for the spacecraft positioned at L1.}
      \label{fig:Registered_coordinates}
    \end{subfigure}
    \hfill
    \caption{Coordinate system definitions and transformations from raw NFI (\subref{fig:NFI_L0_coordinates}) and WFI (\subref{fig:WFI_L0_coordinates}) detector frames to the common registered GSE frame (\subref{fig:Registered_coordinates}).}
\end{figure}

Images collected on orbit will not have consistent orientations with respect to real-world coordinates, since the spacecraft will be rotated each day to keep the payload in shadow. Furthermore, the specific readout architecture induces distinct geometric orientations for the two channels. Specifically, the WFI reference frame is rotated $180^{\circ}$ relative to the NFI. Additionally, images from both channels exhibit a reflection (parity flip) with respect to the standard real-world coordinate system. These coordinate transformations are illustrated in Figures \ref{fig:NFI_L0_coordinates} and \ref{fig:WFI_L0_coordinates}. Finally, any misalignment between the boresights of the two channels will introduce a relative translation between images from the two channels. In order to support temporal image stacking for the general scientific community, all images will be registered to a common coordinate system as shown in Figure \ref{fig:Registered_coordinates}. Registration will be limited to correcting for rotation, reflection, and translation alignment - images will not be re-scaled.

Early in the mission, stellar data will be used to measure the transformation between the spacecraft coordinate frame and each camera coordinate frame. This transformation is expected to be temporally stable. The reported spacecraft attitude will be utilized to register each image to a common coordinate frame. Additionally, image-based registration will be employed to center Earth in nadir NFI images. All images will be transformed such that the projected ecliptic normal (Geocentric Solar Ecliptic (GSE)\footnote{The Geocentric Solar Ecliptic coordinate system is centered at Earth with the $x$-axis pointed towards the Sun, the $y$-axis lying in the ecliptic plane and pointing in the direction opposite to Earth's orbital motion (pointing towards dusk), and the $z$-axis completing a right-handed orthogonal system.} $z$-axis) is aligned with the ``up'' axis of the image (anti-rows) and the projected GSE $y$-axis is aligned with the $x$-axis of the image (columns). Nadir images will further be translated such that Earth is centered.

\section{Science Data Processing Pipeline: L0 to L1A}
\label{sec:sci_data_processing_l0_l1A}

The previous sections established the theoretical basis and performance of the individual calibration algorithms. This section now turns to their integration into the operational Science Data Processing Pipeline. This pipeline serves as the automated framework that systematically applies these corrections to the raw telemetry in order to yield calibrated images. See Table \ref{tab:data_product_def_table} for the list of data product definitions. The first stage of this pipeline, the transition from L0 to L1A, focuses only on detector voltage bias and dark-row removal.

All NFI L0 images have dimensions of $1040 \times 1024$ pixels, while all WFI L0 images have dimensions of $520 \times 512$ pixels. The algorithm is as follows:

\begin{enumerate}
    \item {[Detector Voltage Bias Removal]} Remove the detector voltage bias from all L0 images using the algorithm described in Section \ref{sec:bias_frame_algorithm}.
    \item {[Dark Row Removal]} Remove the electrically and optically dark rows from all images. The NFI images are then reduced to $1024 \times 1024$ pixels, while the WFI images are reduced to $512 \times 512$ pixels.
    \item {[Environment Tagging]} Evaluate the average value of all pixels outside the FOV and convert to units of electron counts per second. If the mean electron count rate is at or below the dark current threshold (to be set on-orbit and defaults to the dark current measured prior to launch), then tag the image as a `quiet-time' image. Otherwise, tag the image as a `storm-time' image.
\end{enumerate}

The science data processing from L0 to L1A is validated by using the numerical image simulator to generate a L0 image for each channel using the baseline science filter (LyaN). The NFI image has an integration time of $30$ minutes, while the WFI image has an integration time of $60$ minutes, which corresponds to the baseline image acquisition scheme for nadir-pointed science images on-orbit. It was found that the percent error on the ground-truth L1A image exhibits significantly higher variance in pixels outside the FOV compared to those within the FOV. This behavior is a direct consequence of the lower expected signal [DN] rates in the un-illuminated regions. Within the FOV, the error distribution is centered at 0\% with a standard deviation of $\sim0.023\%$, whereas outside the FOV, the deviation increases to $\sim9\%$. Since non-FOV pixels are masked in the L1B data product, the elevated noise floor in image corners is considered acceptable.

\section{Science Data Processing Pipeline: L1A to L1B}
\label{sec:sci_data_processing_l1A_l1B}

The second stage of the science data processing pipeline transforms L1A images to L1B images. All L1A images, except for blocked images, have an L1B counterpart. See Table \ref{tab:data_product_def_table} for the list of data product definitions. Recall that L1A images are also tagged as `quiet-time' or `storm-time' (see Section \ref{sec:sci_data_processing_l0_l1A}).

\subsection{Algorithm}

Processing quiet-time L1A images only requires the most recent quiet-time blocked image. Processing storm-time radiation L1A images requires the most recent quiet-time blocked image and the most recent storm-time blocked image taken as close in time to the L1A image as possible. The two conditions are dealt with differently. In the quiet-time case:
\begin{enumerate}
    \item {[Anomalous Pixel Removal]} Subtract the quiet-time blocked image from the L1A image. This removes anomalous pixels at the cost of extra noise from dark current.
    \item {[Dark Current Removal]} Subtract the average of the pixels outside of the FOV from the pixels in the FOV. This removes any remaining dark current.
\end{enumerate}
The radiation subtraction step is omitted as the signal contribution from radiation, in [DN], is negligible.

In the storm-time case:
\begin{enumerate}
    \item {[Anomalous Pixel \& Dark Current Removal]} Subtract the quiet-time blocked image from the L1A image.
    \item {[APS Radiation Removal]} Evaluate the average of the pixels outside of the FOV in the L1A image. This is the retrieved mean [DN] contribution from APS radiation in each pixel in the L1A image. Subtract the retrieved mean APS radiation from all pixels in the L1A image.
    \item {[MCP Radiation Removal]} Scale the mean APS radiation rate by the radiation scaling factor. This yields the mean [DN] contribution from MCP radiation in the FOV. Subtract the mean MCP radiation multiplied by the MCP gain flat from the pixels in the FOV of the L1A image.
\end{enumerate}

Once all extra signal, in [DN], resulting from dark and radiation are removed in either quiet-time or storm-time conditions, the following steps are taken to completely remove all instrument effects:
\begin{enumerate}
    \item {[Optical Flat Field Removal]} Divide by the flat-field.
    \item {[Distortion Removal]} Correct for distortion by using the distortion coefficients.
    \item {[Translation and Rotation]} Translate and rotate the current image so that Earth is centered and ecliptic north is upright, with dusk/dawn on right/left.
\end{enumerate}

The results of these operations are denoted L1B images. To validate the full L0 to L1B calibration pipeline, the numerical image simulator was utilized to generate synthetic L0 exposures for both channels. These simulations employed the baseline science filter (LyaN) and adopted integration times of $30$ minutes for NFI and $60$ minutes for WFI, designed to mimic the nominal parameters of planned nadir-pointed science observations. Other images (such as the flat-field set) are also generated as needed to support this test.

It was found that the percent error on the ground-truth L1B image is centered at 0\% with standard deviations of $2.6\%$ within the FOV for the NFI channel and $1.1\%$ within the FOV for the WFI channel. Pixels outside the FOV are not used in any subsequent retrieval or removal algorithm and contain no information about any photon source and are thus removed. The WFI percent error map exhibits localized artifacts (resembling bubbles) attributed to imperfections in the star-masking process during flat-field retrieval. However, given that these features impact a statistically negligible number of pixels, these outliers are deemed negligible and do not materially impact the global validation results.

\section{Conclusion}

The Carruthers Geocorona Observatory successfully launched in September 2025 and began science operations at the Earth-Sun L1 Lagrange point in January 2026. This work presented the end-to-end science data processing pipeline required to transform raw telemetry into instrument-corrected images, detailing algorithms for voltage bias removal, dark current subtraction, anomalous pixel removal, flat-fielding, radiation signal removal, distortion correction, and image registration. Validation via a numerical simulator demonstrates that the calibrated images are unbiased (centered at 0\%) with residual variations limited to $2.6\%$ (NFI) and $1.1\%$ (WFI). These results confirm that the pipeline effectively isolates and removes the instrument signature, providing a high-fidelity baseline for the subsequent removal of photon backgrounds. The algorithms discussed here establish the pre-launch baseline for operational processing of flight data, providing the foundation for the on-orbit scientific analysis.

\subsection*{Acknowledgements}

\begin{itemize}
\item Funding: This work was supported by the NASA Science Mission Directorate, Heliophysics Division through contract 80GSFC21C0038.
\item Conflict of interest/Competing interests:
Not applicable
\item Ethics approval and consent to participate:
Not applicable
\item Consent for publication:
Not applicable
\item Author contribution: Ordered in author list.
\end{itemize}


\bibliographystyle{ieeetr} 
\bibliography{ref}

\end{document}